\pdfoutput=1

\documentclass[aps,pra,reprint,superscriptaddress,longbibliography]{revtex4-1}

\usepackage{amsmath,amssymb}
\usepackage{mathrsfs}
\usepackage{graphicx}
\usepackage{color}
\usepackage[english]{babel}

\newcommand{\cL}{{\mathcal L}}

\newcommand{\sL}{{\mathsf L}}
\newcommand{\sH}{{\mathsf H}}
\newcommand{\ssb}{{\mathsf{b}}}
\newcommand{\eps}{\varepsilon}

\newcommand{\ket}[1]{\left \vert #1 \right \rangle}

\newcommand{\be}{\begin{equation}}
\newcommand{\ee}{\end{equation}}
\newcommand{\bea}{\begin{eqnarray}}
\newcommand{\eea}{\end{eqnarray}}
\newcommand{\dint}[1]{\mathrm{d} #1 ~}

\begin{document}
 
\title{Ro-Translational Cavity Cooling of Dielectric Rods and Disks}

\author{Benjamin A. Stickler}
\affiliation{Faculty of Physics, University of Duisburg-Essen, Lotharstra\ss e 1, 47048 Duisburg, Germany}
\author{Stefan Nimmrichter}
\affiliation{Faculty of Physics, University of Duisburg-Essen, Lotharstra\ss e 1, 47048 Duisburg, Germany}
\affiliation{Centre for Quantum Technologies, National University of Singapore, Singapore}
\author{Lukas Martinetz}
\affiliation{Faculty of Physics, University of Duisburg-Essen, Lotharstra\ss e 1, 47048 Duisburg, Germany}
\author{Stefan Kuhn}
\affiliation{Faculty of Physics, University of Vienna, VCQ, Boltzmanngasse 5, 1090 Vienna, Austria}
\author{Markus Arndt}
\affiliation{Faculty of Physics, University of Vienna, VCQ, Boltzmanngasse 5, 1090 Vienna, Austria}
\author{Klaus Hornberger}
\affiliation{Faculty of Physics, University of Duisburg-Essen, Lotharstra\ss e 1, 47048 Duisburg, Germany}

\begin{abstract}
We study the interaction of dielectric rods and disks with the laser field of a high finesse cavity. The quantum master equation for the coupled particle-cavity dynamics, including Rayleigh scattering, is derived for particle sizes comparable to the laser wavelength. We demonstrate that such anisotropic nanoparticles can be captured from free flight at velocities higher than those required to capture dielectric spheres of the same volume, and that efficient ro-translational cavity cooling into the deep quantum regime is achievable.
\end{abstract}

\maketitle

\section{Introduction}

Laser cooling and controlling the ro-translational degrees of freedom of a nanoparticle in vacuum is a challenging task \cite{Marago2013} with far-reaching implications: Levitated dielectrics in a high-finesse cavity can be used as ultra-sensitive force sensors with atto- or even zeptonewton sensitivity \cite{Marago2008,Olof2012,Moore2014,Ranjit2016}. In addition, preparing nanoscale dielectrics in the deep quantum regime may allow to address fundamental questions, such as the thermalization of a single particle \cite{gieseler2013,millen2014,gieseler2014,MillenNJP2016} or the validity of the quantum superposition principle at high mass scales \cite{Bassi2013,Arndt2014testing}.

In the case of small molecules with a sharp internal transition, the ro-translational motion can be laser cooled by exploiting the methods developed for atoms \cite{Morigi2007,Shuman2010,Yeo2015}. Micron-sized particles in solution and low vacuum can be trapped and manipulated rotationally with optical tweezers and vortex beams \cite{Paterson2001,Bonin2002,Jones2009,Tong2010,Padgett2011,arita2013,Brzobohaty2015,Simpson2016,spesyvtseva2016}. A first step towards controlling the ro-translational state of a nanometer-sized rod in high vacuum was demonstrated only recently \cite{kuhn2015}.

Here, we show that nanoscale rods and disks are excellent candidates for cavity cooling \cite{kiesel2013,asenbaum2013,millen2015}, possibly even into their ro-translational ground state. This is due to a number of fortuitous properties: $(i)$ The anisotropic shape of a dielectric appreciably enhances the effective interaction with the cavity as compared to a sphere of the same volume, and  $(ii)$ for sufficiently red-detuned cavities, efficient cooling takes place for all orientations and positions of the particle. In addition, $(iii)$ their ro-translational motion can be tracked from the scattered light, since the particle's position and orientation is encoded in the polarization and intensity, and $(iv)$ the final temperature is well below the ro-translational level spacing in the trap potential.

The finite extension of the dielectric must be taken into account when describing the interaction between a nanoparticle and the field of a high finesse cavity since the laser intensity varies on the length-scale of the  particles \cite{romero2010,chang2012,pflanzer2012,kuhn2015}. Rather than using numerical techniques \cite{Voshchinnikov1993,Simpson2014} or iterative methods \cite{Cohen1983}, we exploit that the considered particles are thin, allowing us to analytically derive the internal polarization field \cite{schiffer79} and thus the optical potential. Here, it is crucial that we adopt the direction of the internal polarization field from the exact electrostatic solution \cite{dehulst} and therefore correctly account for the anisotropic susceptibility of the particle. Our resulting analytic expressions for the Markovian particle-cavity dynamics and the scattered light intensity provide the theoretical toolbox required to carry out cavity experiments with dielectric nanorods and nanodisks in high vacuum.

We start by deriving the optical potential of a dielectric rod or disk in a standing wave cavity mode, which enters the Markovian master equation for the combined state of particle and field mode. The cooling rate due to the retarded back-action of the light field is then extracted from this equation. It also determines the threshold velocity for capturing a particle in free flight. We account for recoil heating of the particle by light scattering into the vacuum modes by deriving the orientation-dependent Rayleigh scattering operators, and determine the final temperature of a deeply trapped particle.

\section{Cavity induced potential}

\begin{figure}
 \centering
 \includegraphics[width = 85mm]{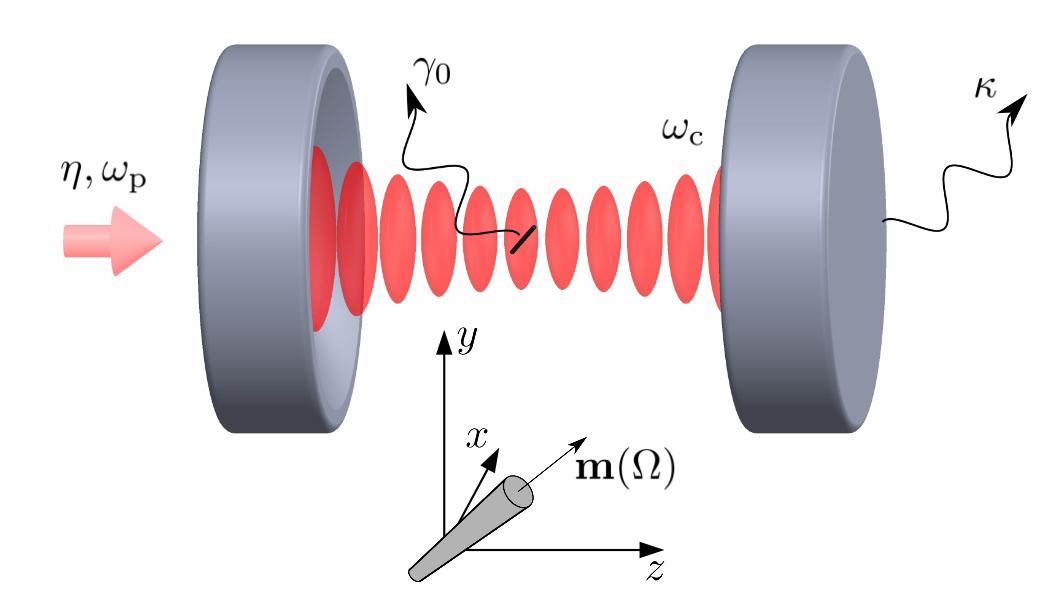}
 \caption{(Color online) A thin dielectric rod or disks traverses the laser field of a high finesse cavity with resonance frequency $\omega_{\rm c}$, driven with pump rate $\eta$ and pump frequency $\omega_{\rm p}$. The cavity detuning is $\Delta = \omega_{\rm p} - \omega_{\rm c}$ and $\kappa$ and $\gamma_0$ are the cavity linewidth and the Rayleigh scattering rate, respectively. The orientation vector of the dielectric's symmetry axis is denoted by ${\bf m}(\Omega)$.} \label{fig:schem}
\end{figure}

We consider a thin dielectric rod or disk, modeled as a cylinder of length $\ell$, radius $a$, and mass $M$ with moment of inertia perpendicular to the symmetry axis, $I_{\rm r} = M \ell^2 / 12$ or $I_{\rm d} = M a^2 / 4$, respectively. The particle propagates through the field of a standing wave Gaussian cavity mode of waist $w_0$ and wavelength $2 \pi / k$, see Fig. \ref{fig:schem}. The cavity mode is driven by a pump laser of angular frequency $\omega_{\rm p}$ and power $P_{\rm p} = \hbar \omega_{\rm p}  \eta^2 / 2 \kappa$ where $\eta$ is the pump rate \cite{ritschnjp2009}. Choosing the $z$-axis along the cavity axis and the $x$-axis in direction of the cavity mode polarization, the local cavity field at position ${\bf r}' = (x',y',z')$ reads ${\bf E}_{\rm in} = \sqrt{2 \hbar \omega_{\rm p} / \eps_0 V_{\rm c}} b f({\bf r}') \cos(k z') {\bf e}_x$ where $f({\bf r}')$ is the Gaussian envelope with waist $w_0$ and $V_{\rm c}$ and $b$ denote the mode volume and the 
dimensionless field amplitude, respectively.

Evaluating the optical force and torque exerted by the laser field on the dielectric requires knowledge of the macroscopic polarization field ${\bf P}$ inside the particle \cite{jackson1999,barnett06}. In general, the internal field must be determined numerically if the particle's extension is comparable to the laser wavelength \cite{Simpson2014}. However, in the present case one can exploit that the particle is very thin in at least one direction in order to derive an approximate analytic expression for the polarization field (generalized Rayleigh-Gans approximation) \cite{schiffer79}. Specifically, this is possible for rods and disks of (real) dielectric permittivity $\eps_{\rm r}$ since $\pi k^2 a^2 (\eps_{\rm r} - 1) \ll 1$ or $k \ell (\eps_{\rm r} - 1) \ll 1$, respectively.

The resulting internal polarization field acquires the position dependence of the external field but it is rotated according to the dielectric's susceptibility tensor. The internal field is the exact solution to Maxwell's equations in the limit of infinitesimally thin particles and it provides a good approximation for the present scenario \cite{kuhn2015}. In particular, the field accounts correctly for the anisotropic susceptibility of the nanoparticles because its polarization is obtained by solving the corresponding electrostatic problem \cite{schiffer79}. In the case of rods, the components of the susceptibility tensor perpendicular and orthogonal to the symmetry axis are $\chi^{\rm r}_\| = \eps_{\rm r} - 1$ and $\chi^{\rm r}_{\bot} = 2 ( \eps_{\rm r} - 1)/(\eps_{\rm r} + 1)$ while $\chi^{\rm d}_\| = (\eps_{\rm r} - 1) / \eps_{\rm r}$ and $\chi^{\rm d}_\bot = \eps_{\rm r} - 1$ for disks \cite{dehulst}. We denote the anisotropy by $\Delta \chi = \chi_\| - \chi_\bot$ and the maximal value by  $\chi_{\rm m} 
= \eps_{\rm r} - 1$. Note that the orientationally averaged susceptibility of a dielectric rod or disk thus exceeds the susceptibility of a dielectric sphere. With the above notation, the polarization field is proportional to $\chi_\bot {\bf e}_x + \Delta \chi ({\bf e}_x \cdot {\bf m} ) {\bf m}$ where ${\bf m}$ is the direction of the dielectric's symmetry axis. This dependence on the dielectric's orientation is familiar from anisotropically polarizable point-like particles \cite{stapelfeldt03}.

The particle-cavity interaction potential can now be calculated by integrating the potential energy density derived from cavity perturbation theory \cite{pozar1997,chang2010,pflanzer2012} averaged over one optical cycle, $-{\bf P} \cdot {\bf E}^*_{\rm in} /4$, or, equivalently, from the averaged force- and torque densities \cite{barnett06}. Denoting the center-of-mass position of the dielectric by ${\bf r}$ and its orientation by ${\bf m}(\Omega)$, with $\Omega = (\alpha,\beta,\gamma)$ the Euler angles in the $z$-$y'$-$z''$ convention, we find the optical potential as
\bea \label{eq:pot}
 V({\bf r},\Omega) & = & \hbar U_0 \vert b  \vert^2 f^2({\bf r}) \left [ \frac{\chi_\bot}{\chi_{\rm m}} + \frac{\Delta \chi}{\chi_{\rm m}} ({\bf m}(\Omega) \cdot {\bf e}_x)^2 \right ] \notag \\
 && \times  \left [ \frac{1}{2} + \frac{1}{2} \cos(2 k z) S({\bf m}(\Omega), {\bf e}_z) \right ].
\eea
Here, $U_0 = - \omega_{\rm p} \chi_{\rm m} V_0 / 2 V_{\rm c}$ denotes the coupling frequency with $V_0$ the dielectric's volume. The orientation-dependent shape function $S({\bf m},{\bf n})$ accounts for the particle's finite extension and reads for rods and disks, respectively, as
\bea \label{eq:sfunc}
S_{\rm r}({\bf m},{\bf n}) & = & \frac{\sin \left (k \ell {\bf m} \cdot {\bf n} \right )}{k \ell {\bf m} \cdot {\bf n}}, \notag \\
S_{\rm d}({\bf m},{\bf n}) & =  & \frac{J_1 ( 2 k a \vert {\bf m} \times {\bf n} \vert)}{k a \vert{\bf m} \times  {\bf n} \vert},
\eea
where $J_1(\cdot)$ denotes a Bessel function of the first kind. Both functions \eqref{eq:sfunc} take on their maximum value if their arguments vanish and, thus, it can be seen from \eqref{eq:pot} that rods tend to align with the field polarization, ${\bf m} = {\bf e}_x$, while disks align with the cavity axis, ${\bf m} = {\bf e}_z$. In the limit of small dielectrics, $k\ell \ll 1$ or $ka \ll 1$, the potential of an anisotropic point-like particle in a standing wave laser field \cite{stapelfeldt03} is recovered and the potential \eqref{eq:pot} becomes proportional to the local laser intensity.

\section{Master equation for dielectric and cavity}

A dielectric particle moving through the cavity modifies the laser intensity by effectively shifting the cavity resonance frequency $\omega_{\rm c}$ as well as enhancing the cavity loss rate. While this retarded reaction of the cavity on the dielectric's motion can cool the particle's motional state, Rayleigh scattering of cavity photons off the dielectric leads to recoil heating. 
Compared to other decoherence mechanisms \cite{chang2010}, recoil heating dominates for deeply trapped particles and thus determines the steady state temperature. 
The coupled particle-cavity dynamics can be described with the help of a Markovian quantum master equation for the total state operator $\rho$. This equation can be derived by coupling the particle-cavity system to the infinite bath of empty vacuum modes, which are then traced out in the Born-Markov approximation \cite{Agarwal}. Here, scattering enhanced coupling between vacuum modes \cite{pflanzer2012} can be safely neglected since we consider very thin particles, for which the 
scattering rate is sufficiently low. Denoting by $\hat {\bf r}$ and $\hat \Omega$ the ro-translational coordinate operators and by $\ssb$ the cavity field operator, one obtains
\bea \label{eq:master}
\partial_t \rho & = & - \frac{i}{\hbar} \left [ \sH_{\rm p},\rho \right ] - i U_0 \left [ v(\hat {\bf r}, \hat \Omega) \ssb^\dagger \ssb, \rho \right ] + \cL_{\rm c}\rho \notag \\
 && + \gamma_0 \sum_{s = 1,2} \int_{S_2} \frac{\mathrm{d}^2 {\bf n}}{4 \pi} \left [ \sL_{{\bf n}s} \rho \sL_{{\bf n}s}^\dagger - \frac{1}{2} \left \{ \sL_{{\bf n}s}^\dagger \sL_{{\bf n}s}, \rho \right \} \right ],
\eea
where $\sH_{\rm p}$ is the free particle Hamiltonian and $\cL_{\rm c}\rho$ describes the unperturbed cavity dynamics,
\be
\cL_{\rm c}\rho = i \left [ \Delta \ssb^\dagger \ssb + i \eta (\ssb - \ssb^\dagger), \rho \right ] + \kappa \left ( 2 \ssb \rho \ssb^\dagger - \{ \ssb^\dagger \ssb, \rho \} \right ),
\ee
with $v({\bf r},\Omega) = V({\bf r},\Omega) / \hbar U_0 \vert b \vert^2$ the field-independent part of the interaction potential \eqref{eq:pot}. The position- and orientation-dependent Lindblad operators of Rayleigh scattering into polarization direction $\boldsymbol{\epsilon}_{{\bf n}s}$ with scattering rate $\gamma_0 = c \chi_{\rm m}^2 V_0^2 k^4/ 6 \pi V_{\rm c}$, are $\sL_{{\bf n}s} = \ssb A_{{\bf n}s}[\hat {\bf r},{\bf m} (\hat \Omega)]$ with
\begin{widetext}
\be \label{eq:lns}
A_{{\bf n}s}({\bf r},{\bf m}) = \sqrt{\frac{3}{8}} f({\bf r}) \boldsymbol{\epsilon}_{{\bf n}s} \cdot \left [ \frac{\chi_\bot}{\chi_{\rm m}} {\bf e}_x + \frac{\Delta \chi}{\chi_{\rm m}} ( {\bf m} \cdot {\bf e}_x) {\bf m} \right ] e^{-i k {\bf n}\cdot {\bf r}}\left ( e^{i k z} S[{\bf m} , ({\bf e}_z - {\bf n})/2] + e^{-i k z} S[{\bf m}, ({\bf e}_z + {\bf n})/2] \right ).
\ee
\end{widetext}
They are diagonal in the ro-translational degrees of freedom and thus tend to localize the state in position and orientation. While the operator \eqref{eq:lns} transfers a superposition of the momentum kicks $\hbar k ({\bf n} \pm {\bf e}_z)$ on the dielectric's center of mass, its action on the orientational degrees of freedom is more intricate because the angle operators $\hat \Omega$ are not the generators of angular momentum kicks but of the canonically conjugate momentum translations \cite{Edmonds1996,Brink2002}. Thus, an eigenstate $\ket{m_\alpha,m_\beta,m_\gamma}$ of the canonical momentum operator $\hat p_\Omega$ conjugate to the Euler angles $\Omega$ transforms under the action of \eqref{eq:lns} into a superposition of shifted eigenstates, each weighted with the corresponding Fourier coefficient of the Lindblad operator. Nevertheless, it will turn out that the Rayleigh scattering dissipator describes diffusive ro-translational motion for deeply trapped nanoparticles. In addition, in the limit of small, isotropically polarizable point particles \cite{chang2010,romero2010} the operators \eqref{eq:lns} turn proportional to the local standing wave profile.
 
\section{Scattered light intensity}

The light scattered by the dielectric can be utilized to track its ro-translational motion \cite{kuhn2015}. Specifically, a detector at position $R {\bf n}$ measures the intensity
\be
I_{\bf n}({\bf r},\Omega) =\frac{\hbar \omega_{\rm p} \gamma_0 \vert b \vert^2}{4 \pi R^2} \sum_{s = 1,2} \vert A_{{\bf n}s}[{\bf r},{\bf m}(\Omega)] \vert^2
\ee
as derived from the electric field integral equation \cite{schiffer79} in the far-field limit. The polarization of the scattered field depends only on the dielectric orientation since it is orthogonal to the direction of its internal polarization field.

\section{Equations of motion}

In order to assess under which conditions cavity cooling and trapping of the dielectric is possible, we determine the local cooling rate from the classical equations of motion. Denoting by $({\bf p}, p_\Omega)$ the canonically conjugate momentum coordinates, the classical equations of motion are obtained from the master equation \eqref{eq:master} by replacing all operators by their expectation values,
\begin{subequations} \label{eq:dyn}
\bea
\dot{b} & = & i \left ( \Delta - U_0 v \right ) b -  \left (\kappa + \frac{ \gamma_{\rm sc}}{2} \right ) b + \eta,  \label{eq:cav} \\
 \dot{\Omega} & = & \partial_{p_\Omega} H_{\rm p}, \quad \text{and} \quad \dot{{\bf r}} = \partial_{\bf p} H_{\rm p},\label{eq:partx} \\
\dot{p}_\Omega & = & - \partial_{\Omega} V + \hbar \gamma_0 \vert b \vert^2 \sum_{s = 1,2} \int_{S_2} \frac{\mathrm{d}^2{\bf n}}{4 \pi} \mathrm{Im} \left ( A_{{\bf n}s}^* \partial_{\Omega} A_{{\bf n}s} \right ),\notag  \\
\text{and} && \notag \\
\dot{{\bf p}} & = & - \partial_{\bf r} V + \hbar \gamma_0 \vert b \vert^2 \sum_{s = 1,2} \int_{S_2} \frac{\mathrm{d}^2{\bf n}}{4 \pi} \mathrm{Im} \left ( A_{{\bf n}s}^* \partial_{\bf r} A_{{\bf n}s} \right ), \label{eq:partp}
\eea
\end{subequations}
where we introduced the total scattering rate as a function of position and orientation,
\be \label{eq:scattrate}
\gamma_{\rm sc}({\bf r},\Omega) = \gamma_0 \sum_{s=1,2} \int_{S_2} \frac{\dint{{}^2{\bf n}}}{4 \pi} \left \vert A_{{\bf n}s}[{\bf r},{\bf m}(\Omega)] \right \vert^2.
\ee

Equation \eqref{eq:cav} describes the retarded reaction of the light field on the particle dynamics. The equations of motion of the particle \eqref{eq:partx} and \eqref{eq:partp} contain the conservative optical potential \eqref{eq:pot} as well as the non-conservative radiation pressure due to Rayleigh scattering \eqref{eq:lns}. This contribution vanishes close to the minimum of the potential \eqref{eq:pot}; it vanishes everywhere for isotropic point particles.

The particle-cavity equations \eqref{eq:dyn} must be solved numerically in general. However, if the particle is not yet deeply trapped, we can neglect Rayleigh scattering. Assuming further that the particle moves sufficiently slowly, such that the cavity reacts nearly instantaneously, we expand the cavity amplitude to first order in all velocities and angular momenta. Thus, we obtain non-conservative equations describing the dissipative dynamics of the nanoparticle. In general, the different degrees of freedom are strongly coupled and exchange energy, such that it is not useful to define a friction rate for the individual coordinates. Nevertheless, by adapting Liouville's theorem one can calculate the rate at which an infinitesimal phase space volume centered at $({\bf r},\Omega)$ expands or contracts \cite{ezra},
\bea \label{eq:rate}
\Gamma({\bf r},\Omega) & = &  \frac{4 \hbar \kappa \eta^2 U_0^2\left [ \Delta - U_0 v({\bf r}, \Omega) \right ] }{\left ( \kappa^2 + [ \Delta - U_0 v({\bf r},\Omega)]^2 \right )^3} \notag \\
 &&\!\!\!\!\!\!\!\!\!\! \times \left ( \frac{[\partial_{\bf r} v({\bf r},\Omega)]^2}{M} + \frac{[\partial_\alpha v({\bf r},\Omega)]^2}{I \sin^2 \beta} + \frac{[\partial_\beta v({\bf r},\Omega)]^2}{I} \right ).
\eea
Notably, this rate is everywhere negative if the laser is sufficiently far red-detuned, {\it i.e.} for $\Delta < U_0$, and thus cooling occurs irrespective of the particle's position and orientation. This generalizes the results obtained for cavity cooling of point particles \cite{hechenblaikner1998,RitschRev2013}. The cooling rate \eqref{eq:rate} vanishes if the particle reaches the mechanical equilibrium, where the partial derivatives of $v({\bf r},\Omega)$ are zero. This could be circumvented by resorting to well established techniques such as two-mode \cite{chang2010,habraken2013} or feedback-cooling \cite{Hohberger2004,li2011,Gieseler2012,vovrosh2016}.

\begin{figure}[t!]
 \centering
 \includegraphics[width= 90mm]{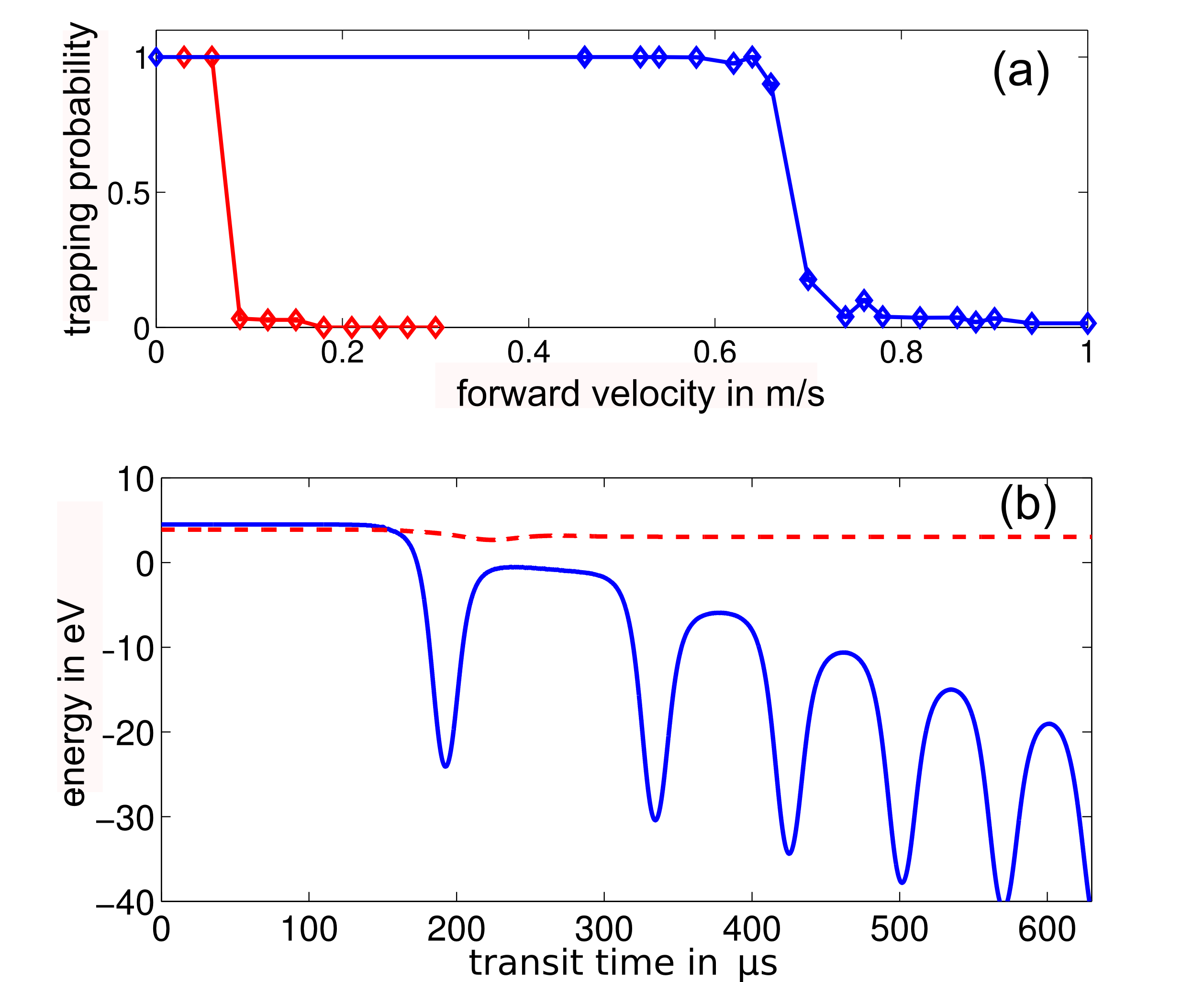}
 \caption{(Color online) (a) Trapping probability as function of the forward velocity for a silicon rod (blue; $\ell = 800$ nm, $a = 25$ nm) and a sphere (red; $R \simeq 72$ nm) of the same volume. The nanoparticle is launched towards the cavity with forward velocity $v_x$. In order to obtain the trapping probability we solve the classical equations of motion \eqref{eq:dyn} for several thousand initial conditions for each $v_x$. The transverse velocity $v_z$ is uniformly distributed within $5$ \% of the forward velocity $v_x$ assuming a collimated beam. We neglect the weak $y$-dependence, setting $v_y = 0$ and $y = 0$, and assume the orientational degrees of freedom to be micro-canonically distributed with a total rotation frequency of $1$ MHz. For the cavity, we chose the following realizable parameters: $\lambda=1.56\,\mu$m, $\kappa = 0.78$ MHz, $P_{\rm in} = 10$ mW,  $\Delta = -1.2 \kappa$, $F = 330000$. In (b) we show the total energy for a sample trajectory with $v_x = 0.5$ m/s and $v_z = -0.3$ m/s. The 
dielectric rod (blue) 
is captured and cooled while a sphere (red dashed) of the same volume traverses the cavity almost unaffected. Negative energies indicate that the particle is captured.} \label{fig:trapping}
\end{figure}

The facts that cooling occurs almost everywhere for $\Delta < U_0$ and that the effective interaction strength is enhanced due to the anisotropy have a remarkable consequence: They facilitate trapping of rods and disks at velocities at which corresponding spheres would traverse the cavity almost unaffected. This is demonstrated in Fig. \ref{fig:trapping}, which shows the trapping probability as a function of the forward velocity for dielectric rods and spheres of the same volume together with the total energy along one sample trajectory.

\section{Cooling limit}

Recoil heating due to Rayleigh scattering prevents the dielectric from being cooled to its absolute ground state. To determine the steady-state temperature, we study the quantum master equation \eqref{eq:master} in the case that the particle is already deeply trapped. For simplicity, we consider a moderately sized particle, $k \ell \simeq 1$ or $ka \simeq 1$, for which the interaction potential \eqref{eq:pot} can be approximated harmonically around its single minimum implying that the ro-translational degrees of freedom are only coupled via the cavity field. (If a linear coupling between the different degrees of freedom is required by a specific protocol this can be achieved by illumination with a second laser.)

For a dielectric rod aligned with the field polarization axis, ${\bf m}(\Omega_0) = {\bf e}_x$, at a position of maximum laser intensity, ${\bf r}_0 = 0$, the harmonic frequencies are
\bea\label{eq:freq}
\omega^{\rm r}_{z} & = & \sqrt{\frac{2 \hbar \vert U_0 \vert \vert b_0 \vert^2 k^2 }{M}}, \text{  } \omega^{\rm r}_{\alpha} = \sqrt{\frac{2 \hbar \vert U_0 \vert \vert b_0 \vert^2  \Delta \chi}{I \chi_\|}}, \notag \\
\omega^{\rm r}_{\beta} & = & \sqrt{\frac{2 \hbar \vert U_0 \vert \vert b_0 \vert^2}{I} \left ( \frac{\Delta \chi}{\chi_\|} + \frac{(k \ell)^2}{12} \right )}.
\eea
In a similar fashion, a deeply trapped disk has its surface aligned with the cavity axis, ${\bf m}(\Omega_0) = {\bf e}_z$, at a position of maximum laser intensity and the frequencies are $\omega_z^{\rm d}= \omega^{\rm d}_\beta = \omega^{\rm r}_z$ together with $\omega^{\rm d}_{\alpha} = 0$. Here $b_0 = \eta/ (\kappa + \gamma_{\rm sc}^{0} / 2 + i [\Delta - U_0])$ denotes the steady-state cavity amplitude in the potential minimum with $\gamma_{\rm sc}^{0} = \gamma_{\rm sc}({\bf r}_0,\Omega_0)$ the corresponding Rayleigh scattering rate. The frequencies in the transversal direction $(x,y)$ can be safely neglected because the laser waist $w_0$ is typically much larger than the wavelength. The frequencies \eqref{eq:freq} are of the same order of magnitude exceeding the cavity linewidth $\kappa$. Note that the trapping frequency $\omega_z$ is determined by the maximum susceptibility $\chi_{\rm m} = \eps_{\rm r} - 1$ rather than 
by the average value.

Similarly, we expand the Lindblad operators harmonically around the potential minimum. A straightforward calculation demonstrates that the remaining Lindblad operators are linear in the field operator $\ssb$ as well as in the position operators $\hat z$ and $\hat \Omega$ and thus they describe diffusive motion. One can determine an approximate expression for the steady-state temperature from the resulting relations between steady-state operator expectation values. Defining temperature as the energy difference between the steady-state energy expectation value and the energy minimum divided by Boltzmann's constant, its steady-state value is given by the recoil limit
\bea \label{eq:temp}
T_\nu & = & \frac{\gamma_0 \hbar^2 \vert b_0 \vert^2}{2 M_\nu (\kappa + \gamma_{\rm sc}^0/2) k_{\rm B}} \sum_{s =1,2} \int_{S_2} \frac{\mathrm{d}^2 {\bf n}}{4 \pi} \vert \partial_\nu A_{{\bf n}s}^0 \vert^2,
\eea
for $\nu = z,\alpha,\beta$ with $M_z = M$, $M_{\alpha,\beta} = I$, and $\partial_\nu A_{\bf n}^0$ the derivative in direction $\nu$ of $A_{\bf n}$ evaluated at the potential minimum. In particular, for small particles, $k \ell \ll 1$, one has $T_z \simeq \gamma_0  \hbar^2 k^2 \vert b_0 \vert^2/ 5 M  \kappa_{\rm eff}  k_{\rm B}$ and $T_{\alpha,\beta} \simeq \gamma_0 \hbar^2 \Delta \chi^2 \vert b_0 \vert^2 / 2 I \chi_{\rm m}^2 \kappa_{\rm eff} k_{\rm B}$ where  $\kappa_{\rm eff} = \kappa + \gamma_0 / 2$.

As an example we consider the silicon rods from Fig. \ref{fig:trapping}. They are strongly coupled to the cavity, $\vert U_0 \vert / \kappa \simeq 1.1$, yielding the final temperature $T_z \simeq 14$ $\mu$K, which corresponds to a mean occupation number in the harmonic potential of $n_z \simeq 0.16$. In a similar fashion, we obtain for the rotational degrees of freedom $T_{\alpha} \simeq 31$ $\mu$K ($n_{\alpha} \simeq 0.34$) and $T_\beta \simeq 29$ $\mu$K ($n_\beta \simeq 0.23$). This demonstrates that reaching the ro-translational ground state should indeed be possible.

\section{Conclusions}

Our findings open the door for numerous experiments and applications: The control gained over center of mass and the orientational degrees of freedom can be used for inertial sensing \cite{Marago2008}. By monitoring the scattered light intensity one can track the dynamical impact induced by a background gas, allowing the direct observation of the thermalization of isolated orientational degrees of freedom \cite{MillenNJP2016} or, by using a directed beam of ultra-cold atoms, one can measure the scattering cross section, thus probing the dispersion interaction of nanoscale dielectrics \cite{Emig2007}.

Ground-state cooling of the nanoparticle would comprise a first step towards optomechanical experiments involving both the center-of-mass and the orientational degrees of freedom \cite{bhattacharya2007,Shi2015}. Such deeply trapped particles can be used as point sources for orientation-dependent interference experiments \cite{shore2015,Stickler2015b} by switching off rapidly the cavity \cite{romeroisart2011,Bateman2014}. If the laser intensity is reduced adiabatically \cite{barker2010}, on the other hand, a free quantum state of low kinetic and rotational energy can be generated. Finally, aligning many anisotropic particles in a single cavity mode might give rise to novel phenomena, such as a non-polar version of a gas of interacting dipoles \cite{Moses2015}, where  synchronization of the dielectric's motion may be observable \cite{Simpson2016}.

\acknowledgments{S. K. and M. A. acknowledge financial support from the Austrian Science Fund (FWF): P 27297 and  W1210-3. We thank James Millen for stimulating discussions.}


\end{document}